%% file: main.tex
\documentclass{article}

\usepackage{microtype}
\usepackage{graphicx}
\usepackage{booktabs} 
\usepackage{pgfplots}
\usepackage{pgfplotstable}
\usepackage{hyperref}
\usepackage{tabularray}

\usepackage{hyperref}

\usepackage[accepted]{icml2024}

\usepackage{amsmath}
\usepackage{amssymb}
\usepackage{mathtools}
\usepackage{amsthm}
\usepackage{xspace}
\usepackage{multirow}
\usepackage{tcolorbox}
\usepackage{adjustbox}
\usepackage{makecell}
\usepackage{subfig}
\usepackage[capitalize,noabbrev]{cleveref}

\usepackage{balance}

\theoremstyle{plain}

\theoremstyle{definition}

\theoremstyle{remark}

\pgfplotsset{compat=1.17}

\usepackage[textsize=tiny]{todonotes}

\newcommand{\tool}{Deep-Bench\xspace}
\newcommand*\circled[1]{\tikz[baseline=(char.base)]{
            \node[shape=circle,draw,inner sep=2pt] (char) {#1};}}

\icmltitlerunning{\tool: Deep Learning Benchmark Dataset for Code Generation}

\begin{document}

\input{macros}

\twocolumn[
    \icmltitle{\tool: Deep Learning Benchmark Dataset for Code Generation}



\icmlsetsymbol{equal}{*}

\begin{icmlauthorlist}
\icmlauthor{Alireza Daghighfarsoodeh}{yyy}
\icmlauthor{Chung-Yu Wang}{yyy}
\icmlauthor{Hamed Taherkhani}{yyy}
\icmlauthor{Melika Sepidband}{yyy}
\icmlauthor{Mohammad Abdollahi}{yyy}
\icmlauthor{Hadi Hemmati}{yyy}
\icmlauthor{Hung Viet Pham}{yyy}
\end{icmlauthorlist}

\icmlaffiliation{yyy}{Department of EECS, York University, Toronto, Canada}

    \icmlkeywords{Benchmark, Large Language Model, ICML}
    
\vskip 0.3in
]

\begin{abstract}
\input{0_abstract}
\end{abstract}

\input{1_intro}

\input{2_related}

\input{3_construction}

\input{4_eval}
\input{5_quantitative}
\input{6_qualitative}
\input{7_limit}
\input{8_conclusion}

\newpage

\nocite{langley00}
\balance

\bibliographystyle{icml2024}

\input{main.bbl}
\newpage

\input{10_appendix}

\end{document}

%% file: 0_abstract.tex
Deep learning (DL) has revolutionized 
areas such as computer vision, natural language processing, and more. However, developing DL systems is challenging due to
the complexity of DL workflows. Large Language Models (LLMs), such as GPT, Claude, Llama, Mistral, etc., have emerged as promising tools to assist in DL code generation, offering potential solutions to these challenges. Despite this, existing benchmarks such as DS-1000 are limited, as they primarily focus on small DL code snippets related to pre/post-processing tasks and lack a comprehensive coverage of the full DL pipeline, including different DL phases and input data types.

To address this, we introduce \tool, a novel benchmark dataset designed for function-level DL code generation. \tool categorizes DL problems based on three key aspects: phases such as pre-processing, model construction, and training; tasks, including classification, regression, and recommendation; and input data types such as tabular, image, and text. 

GPT-4o---the state-of-the-art LLM---achieved 31\% accuracy on \tool, significantly lower than its 60\% on DS-1000. We observed similar difficulty for other LLMs (e.g., 28\% vs. 54\% for Claude, 21\% vs. 41\% for LLaMA, and 15\% vs. 20\% for Mistral). This result underscores \tool's greater complexity. We also construct a taxonomy of issues and bugs found in LLM-generated DL code, which highlights the distinct challenges that LLMs face when generating DL code compared to general code.

Furthermore, our analysis also reveals substantial performance variations across categories, with differences of up to 7\% among phases and 37\% among tasks. These disparities suggest that \tool offers valuable insights into the LLMs' performance and areas for potential improvement in the DL domain. 


%% file: 1_intro.tex
\section{Introduction}
In recent years, machine learning (ML) and deep learning (DL) have advanced significantly and
have been integrated into various fields~\cite{hordri2016deep, kamilaris2018deep, gamboa2017deep}. DL coding has its challenges~\cite{arpteg2018software}, and because of its widespread use, many DL systems are developed by domain experts who are often not software developers~\cite{park2021facilitating, singaravel2020explaining}, which amplifies the problems even more.



Recently, with the rise of Large Language Models (LLMs) such as ChatGPT, LLMs are considered among the best solutions for coding tasks~\cite{wang2021codet5, feng2020codebert, achiam2023gpt}.
As shown in Table~\ref{tabs:other_benchmarks}, numerous code generation datasets~\cite{hendrycks2measuring,austin2021program,agashe-etal-2019-juice, lu2021codexglue, yu2018spider,du2024evaluating, zhuo2024bigcodebench} benchmark LLMs' code generation capabilities.
However, DS-1000~\cite{lai2023ds} is the only dataset offering a limited set of DL-specific code generation samples. Specifically, DS-1000 provides small code snippets, typically just a few lines, primarily focused on pre/post-processing tasks. Furthermore, they do not include categorizations such as DL phases (e.g., pre/post-processing, model construction, training, inference, evaluation) or input types (e.g., tabular, image, text), which could provide valuable insights for advancing DL code generation research.



To address this gap, we introduce \tool, a novel dataset designed to benchmark DL code generation at a functional level. \tool includes the code generation prompt, the ground-truth code at the function level, an extensive set of unit tests, and three categorizations (DL phases, ML tasks, and input data types). Unlike DS-1000, \tool provides a more comprehensive set of entries that cover the full DL pipeline, encompassing code examples for various machine learning tasks. Additionally, it includes DL functions with diverse input data types, including tabular, image, and text, making it a more diverse benchmark for DL code generation evaluation.
Specifically, each entry of \tool dataset is categorized based on these three categories which allow a more in-depth evaluation of future DL code generation techniques:
(1)\textit{The DL/ML pipeline stages} (consist of pre/post-processing, model construction, training, inference, and evaluation), (2)\textit{The DL/ML tasks} (consist of classification, object detection, image segmentation, time-series prediction, recommendation, and regression), and (3)\textit{The input data types} (consist of text, image, and array).

Our study found that while GPT-4o demonstrated strong performance in various code generation tasks, it achieved only 31\% accuracy on \tool, significantly lower than the 60\% reported for DS-1000. We observed similar difficulty for other LLMs (e.g., 28\% vs. 54\% for Claude, 21\% vs. 41\% for LLaMA, and 15\% vs. 20\%  for Mistral) which indicates that \tool contains more challenging data points. Furthermore, the difficulty of generating code varies significantly across categories. For example, GPT-4o reaches an accuracy of 33\% for pre/post-processing tasks but only 26\% for model construction. Additionally, the accuracies vary even more among task types, ranging from 58\% for recommendation tasks to 21\% for segmentation tasks. These large gaps in performance across categories highlight the insights that \tool can bring to help improve the LLM DL code generation capability.

To further analyze the root causes of such a poor generation performance, we construct a bug taxonomy of the issues found in the generated DL code. Our qualitative analysis reveals that when compared to LLM-generated general code, LLM-generated DL code exhibits a higher frequency of \textit{deviation from the prompt} issues. Furthermore, \textit{arithmetic and logical errors} emerged as a new and frequently observed category in generated DL code.

\tool's data is available in our public repository\footnote{\href{https://anonymous.4open.science/r/DL-Bench-D65E/}{https://anonymous.4open.science/r/DL-Bench-D65E/}}.

\input{tabs/Others}




%% file: tabs/Others.tex
\begin{table}
\centering
\caption{Code generation benchmarks
(SO-Stack Overflow, EM-exact match, ES-edit similarity, CS-competition specifics).
}
\label{tabs:other_benchmarks}
\resizebox{\linewidth}{!}{

\begin{tabular}{@{}lrllll@{}}
\toprule
\textbf{Benchmark}       & \textbf{Size}         & \textbf{Language}        & \textbf{Level}      & \textbf{Source}      & \textbf{Metrics}         \\
\midrule
Human Eval\cite{chen2021evaluating}      & 164          & Python          & Function          &     Manual        & Pass@k          \\
MBPP\cite{austin2021program}            & 974          & Python          & Function          &       Manual      & Pass@k          \\
APPS\cite{hendrycks2measuring}            & 10000        & Python          & Function          &      Manual       & Pass@k          \\
CoderEval\cite{yu2024codereval}       & 460          & Python-Java     & Function          &     GitHub        & Pass@k          \\
RepoEval\cite{zhang2023repocoder}       & 1600         & Python          & Function         & GitHub          & EM,ES           \\
DS-1000\cite{lai2023ds}         & 1000         & Python          & Statement          & SO          & Pass@k          \\
MLE-Bench\cite{chan2024mle}  & 75           & Python          & Workflow & Kaggle           & CS \\
\midrule
\textbf{\tool} & \textbf{520} & \textbf{Python} & \textbf{Function} & \textbf{GH} & \textbf{Pass@k} \\
\bottomrule
\end{tabular}
}
\end{table}

%% file: 2_related.tex
\section{Related Works}

\textbf{Code Generation for ML software}: Shin et. al.\cite{shin2023good} have explored the effectiveness of neural code generation models on ML tasks, which differ significantly from general programming tasks. A key study evaluated six state-of-the-art models across popular ML libraries, highlighting the need for domain-specific benchmarks and improvements tailored to the complexities of ML. Our work provides such benchmarks specifically for ML and DL software.

Recently, DS-1000\cite{lai2023ds} is the only benchmark designed for data science code generation. It contains 1000 problems sourced from 
StackOverflow with
diverse, practical tasks. Similarly, MLE-bench~\cite{chan2024mle} targets evaluating ML engineering workflows in Kaggle competitions.
However, \tool differs in key aspects: 1) it focuses on ML/DL tasks rather than general data science and MLE, 2) we categorize the data by ML phases, task types, and data types, and 3) our granularity is at the function level rather than script or workflow level. Additionally, unlike the other datasets, our dataset is based on GitHub repositories with real code.

\textbf{Code Generation Benchmarks: } There have been multiple code generation benchmarks as shown in Table~\ref{tabs:other_benchmarks}.
Among them, HumanEval~\cite{chen2021evaluating} is the most popular
with 164 hand-crafted programming problems.
Building on this foundation, AiXBench\cite{hao2022aixbench} extends the evaluation to Java, and MultiPL-E\cite{cassano2022multipl} expands
by supporting 18 different programming languages.
Another prominent benchmark is MBPP~\cite{austin2021program} which offers 974 programming tasks solvable by entry-level programmers.
Additionally, the Spider benchmark~\cite{yu2018spider} provides 10,000 questions paired with 5,000 SQL queries. 
CoderEval~\cite{yu2024codereval} and APPS benchmark~\cite{hendrycks2measuring} assess code generation models in real-world programming scenarios in Python and Java with coding problems ranging from beginner to advanced competitive programming challenges. RepoEval~\cite{zhang2023repocoder}, meanwhile, assesses LLMs for repository-level code generation. All the above-mentioned benchmarks focus on general programming, unlike \tool, which focuses on DL/ML code generation problems.


%% file: 3_construction.tex
\section{Benchmark Construction}
\label{sec:BenchConstr}

\input{figs/pipeline}
\input{figs/codeprompt}

\tool consists of 520 instances of AI and DL data points (filtered from over 2,000 raw data points). The data is curated from 30 GitHub repositories (selected from an initial pool of 160 related repositories).

The construction process of \tool consists of two main phases: The Raw Data Extraction and the Labeling Procedure. The raw data extraction involves six semi-automatic steps. Since \tool is designed to have diverse and realistic code samples, the first step \circled{1} is to construct \tool from code crawled from highly rated GitHub repositories (i.e., with the most stars) filtered using 30 DL-related terms such as ``neural-networks'',  ``pytorch'', ``computer-vision''. We then manually select (step \circled{2}) 160 high quality candidate DL projects 
(i.e., involve the integration of DL and AI-related frameworks, comprehensive test cases, clear and well-written docstrings, and detailed contribution guidelines). We then employed a bespoke utility to extract the test files and then test cases from each repository (step \circled{3} and \circled{4}). By performing static analysis, we were able to track and collect all of the functions under test in step \circled{5} to form the raw data that is the base of \tool. 

Once the raw data is extracted, the labeling procedure starts. To speed up the task of constructing the prompt for each code sample, we utilize LLM (i.e., GPT-4o) as a code-explanation tool\cite{nam2024using} to generate the first prompt candidate for each function under test (step \circled{7}). Four co-authors were then tasked with manually filtering each entry to ensure that each function is highly relevant (i.e., contributes to a DL task such as image recognition, utilizes at least one recognized DL framework, and implements a relatively advanced and sophisticated algorithm).
Finally, we conduct a manual labeling process involving four co-authors (step \circled{9}) to refine the prompt and label each code sample with the appropriate category from our three chosen types of categories: DL pipeline phases, ML task types, and input types.

\subsection{Raw Data Extraction}

This phase consists of six semi-automatic steps that crawl data from GitHub repositories to generate a list of function definitions and their test cases.



\textbf{Repository Selection:} We curated our data from the top 1000 starred DL-related GitHub repositories to include high-quality and widely used DL-related functions.

In step \circled{1}, we filtered GitHub projects with one of 30 DL-related tags such as ``neural-networks'',  ``pytorch'', and ``computer-vision'' (we provided the complete list of tags in our repository). 
Specifically, we select the tags
by collecting from DL and AI-related GitHub repositories and filtering the most relevant ones to get the final 30.

In step \circled{2}, we select 160 most relevant projects for \tool and
retain only projects that: 1) are DL related (i.e., use DL libraries, or perform DL tasks like segmentation or detection), 2) have sufficient test cases (averaging at least three per function), and 3) include thorough documentation, such as source code docstrings or README files. 

\textbf{Function Extraction:} One of the main design choices of \tool is to include a set of reliable and robust test cases for each benchmark entry. This is because programming languages are different from natural languages. Specifically, generated code can fulfill all of the functional requirements but could have a low BLEU score when compared with the ground truth code\cite{tran2019does}. This means that using text similarity metrics such as BLEU score as evaluation metrics is not the best method to evaluate code generation techniques. Instead, test cases (functional and non-functional) passing rate should be used to reliably access a new code generation approach.

In step \circled{3}, we crawled selected repositories for test files using standard test file name patterns such as tests/test\_file\_name.py~\cite{madeja2021automating}. In step \circled{4}, for each test file, we extract test cases using common patterns in Python test suites, such as the @pytest decorator.

Once we identified all test cases, in step \circled{5}, we performed call graph analysis to track and collect all functions under test (excluding third-party function calls). The definitions of each of those functions are then extracted in step \circled{6} to form the bases for our ground-truth code samples.



\subsection{Labeling Procedure}

The labeling procedure involves three semi-automatic steps to generate and refine a prompt and assign categorizations for each entry in our \tool dataset. To determine the best procedure and criteria for our manual process, we perform a small trial run of the manual process on a small sample of the data points. In this trial run, we ask each reviewer to provide feedback on the labeling criteria so that when we start our full run we have
the most comprehensive and accurate manual process possible.

\textbf{Prompt Generation:} 
In step \circled{7}, we utilize two sources of data to create the code generation prompts: 1) the doc-strings provided by developers, which describe the functionality and parameters of the code, and 2) the function definitions themselves, which can be used to generate candidate prompts. Specifically, We take advantage of the function definitions to explain the code, and by combining them with their respective doc-strings (when available), we generate the initial candidate prompt by querying GPT-4o with the template as described in Fig.~\ref{fig:code_prompt}.




However, generated prompts require manual validation to ensure accuracy and relevance. This review process is essential to refine prompts and guarantee quality for subsequent use~\cite{shrivastava2023repository}. We further refine prompts based on the following criteria: (1) contain clear, sufficient information for code generation, (2) specify input and output format, and (3) cover error handling and boundary conditions. More details are in the appendix.

If the prompt does not meet the
mentioned criteria, the annotators propose and agree on changes that 
bring it up to the expected quality.
This reviewing process 
produces prompts that are 
not only 
technically correct 
but also include details essential to
code generation.

\textbf{Data Filtering and Validation:} After compiling all the data (i.e., the ground truth, test cases, and candidate prompts), in step \circled{8}, we manually evaluate each function meticulously, reading and modifying the prompts following a set of criteria. Specifically, we discard general codes (e.g., those for reading text files) that are not DL related.
In this step, the annotators independently assess the
prompt’s clarity, relevance to DL-related tasks, and overall usability 
with the following
criteria: (1) serving key DL tasks, (2) utilization of popular DL frameworks, and (3) algorithms' relevancy and clarity.

\textbf{Labeling:} In step \circled{9}, we assign labels for each data point based on the role of the function in the ML pipeline (e.g., pre/post-processing, model construction), the ML tasks (e.g., classification, regression) it solves, and types of data (e.g., image, text) it operates on. For each data point, three co-authors thoroughly analyze and assign appropriate labels. We use a majority vote to finalize the labels and modify the prompts accordingly. Specifically, we assign the following labels when appropriate to each data point: Stage in the ML pipeline, ML task type, and Input data type.

Once each reviewer completes their assessments, the team meets to discuss any discrepancies and reach a consensus on the final labels. Due to our detailed instructions and guidelines, we achieve a high inter-rater reliability of 0.83 measured by Krippendorff's alpha~\cite{zapf2016measuring}(measures of more than 0.8 indicating strong agreement.

The labeled data is carefully documented, including notes on the decision-making process for transparency and future reference. Instances are organized, with labels to ensure easy retrieval and analysis in later stages of research. To enable easier benchmark utilization (i.e., running test cases), the relevant projects are set up in virtual environments along with appropriate dependencies and ready-to-run testing scripts.

This rigorous review and labeling process ensures that each instance in the dataset is not only relevant and useful but also thoroughly understood and appropriately categorized, contributing to a robust and reliable benchmark. 

%% file: figs/pipeline.tex
\begin{figure*}[t!]
    \centering
    \includegraphics[width=0.9\linewidth]{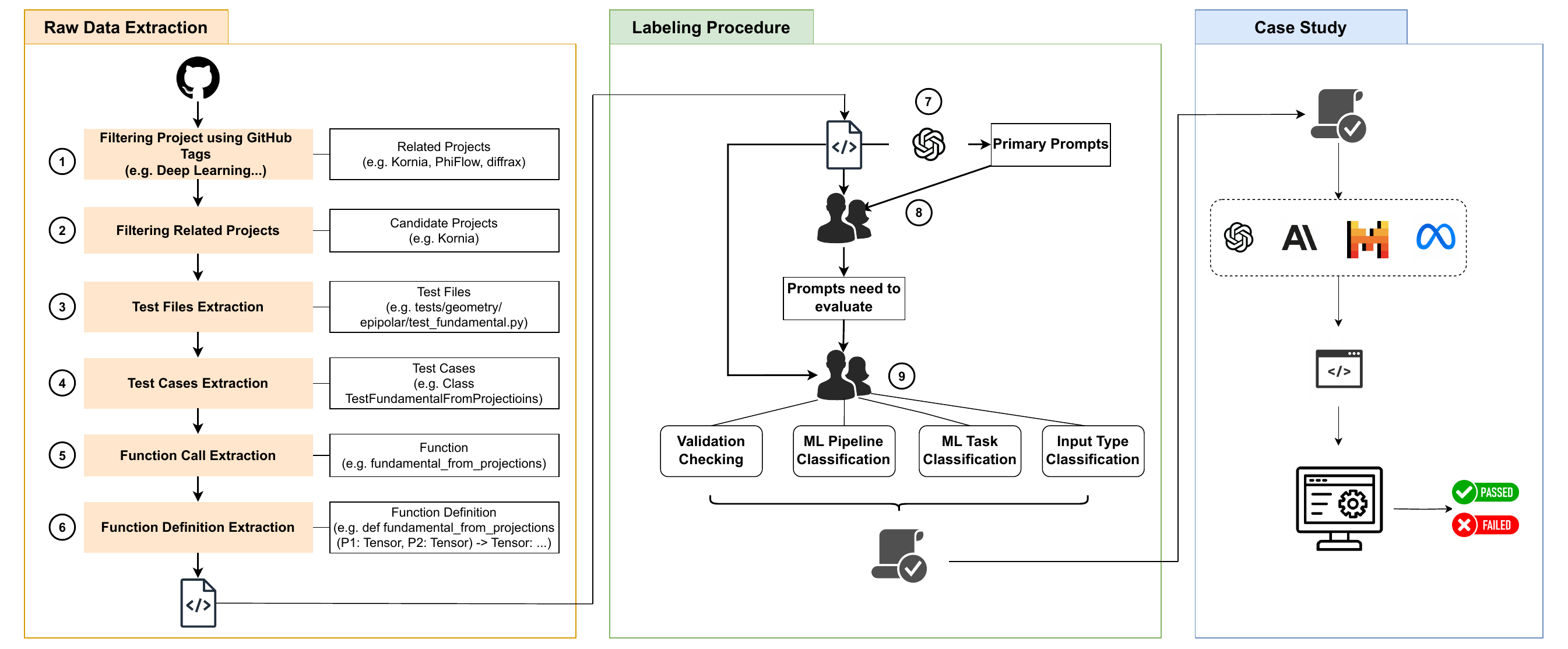}
    \caption{\tool construction procedure
    }
    \label{fig:pipeline}
\end{figure*}

%% file: figs/codeprompt.tex
\begin{figure}
    \centering
    \includegraphics[width=\linewidth]{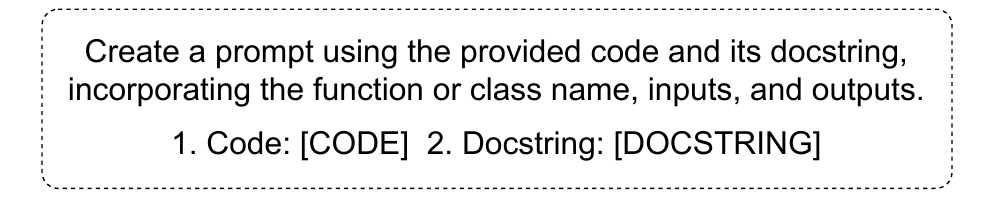}
    \caption{Template of generating prompt from code}
    
    \label{fig:code_prompt}
    \vspace{-10pt}
    
\end{figure}

%% file: 4_eval.tex
\section{Evaluation}

To demonstrate the potential of \tool and the depth of analysis that can be done on \tool, we perform a preliminary study evaluating the performance of state-of-the-art LLM on the \tool benchmark. We choose the top LLMs: 
GPT-4o, 
Claude3.5 sonnet, 
Llama 3.1 70B, and 
Mistral 7B.
The LLMs are evaluated based on the commonly used pass@k, which measures the likelihood that at least one of the k-generated solutions passes all test cases~\cite{lyu2024top}. 


To minimize non-determinism and improve reproducibility, we set the temperature to zero for LLMs \cite{bommasani2021opportunities}. This allowed us to accurately measure the 
quality of the generated code across various DL tasks, offering deeper insights into the LLM's strengths and weaknesses.

For each project, \tool provides 
a
docker image that includes all required dependencies.
These images
ensure that test cases are runnable and the evaluations are
easily replicable on other systems.

%% file: 5_quantitative.tex
\setlength{\parskip}{0pt}

\section{Quantitaive Analysis}
\label{sec:study}

\input{tabs/RQ1_ACC}

\textbf{What are the performances of SOTA LLMs on DL/ML code generation tasks?}
In this analysis, we investigate how the existing ML code generation benchmark (DS-1000) and \tool evaluate SOTA LLMs.
Table~\ref{tab:RQ1_ACC} shows the pass@1 metric of SOTA LLMs on those benchmarks.
To focus on 
demonstrating \tool's ability to evaluate existing LLMs, we intentionally avoided using specialized prompt strategies, opting instead for vanilla prompts to assess the model's baseline performance. However, the use of advanced prompt engineering strategies could yield different results and this will be interesting for future usage of \tool. 

Our analysis underscores that even the most advanced model such as GPT-4o struggles with ML/DL-specific code generation. Specifically, GPT-4o achieves only 60\% and 31\% Pass@1 scores on DS-1000 and \tool respectively. 
Also, when comparing between DS-1000 and \tool, our benchmark is more challenging as the SOTA LLM gets a much lower Pass@1 score, SOTA LLMs in other categories are behind GPT-4o with performance as low as 15\% for the tiny Mistral 7B model on \tool. The overall weak performance of these models highlights the ongoing challenges in generating reliable, executable ML/DL code which supports the need for more benchmarks in this domain.





Overall, GPT-4o's pass@k is 31\%, but to further assess its performance, we calculated the average passing rate across the functions, which was higher at 40\%. This suggests that while only 31\% of functions pass all test cases, many pass some. With additional insights, these partially valid cases could be improved.




\begin{tcolorbox}[boxrule=0.5pt, colback=gray!10,  arc=4pt,left=3pt,right=3pt,top=3pt,bottom=3pt,boxsep=0pt
]
\textbf{Finding 1:} Our evaluation indicates that current LLMs struggle to generate correct, executable code for ML/DL tasks. Although GPT-4o is the strongest among the tested models, it still falls short of meeting practical standards (Pass@k score of only 31\%). 
A deeper analysis is needed (which \tool can provide) to extract insight into improving prompting techniques.
\end{tcolorbox}


\input{tabs/RQ2_ACC}

\textbf{Which stages in the ML pipeline pose a greater challenge for SOTA LLMs?}
We conduct an analysis of the challenges of generating DL code for specific stages in the pipeline (enabled by 
\tool provided categorization)
Table~\ref{tab:RQ2_MLStage} presents the Pass@1 scores that each LLM achieves for code in each 
ML/DL pipeline stage.
Our result shows that the best SOTA LLM---GPT-4o---outperforms all of the other LLMs in all stages. Claude 3.5 sonnet is the closest second, where in the \textit{Inference} category, it is level with GPT-4o. 


\textit{Pre/post Processing} stages often require small but important data transformation tasks, thus such codes are the most available. Hence, 
\tool collected the most data in this category (210/520). 
Code in this category prepares and cleans the input data for the model and formats the model's output. 
Our result shows that LLMs have the most success in generating code for 
the Pre/post Processing stages. 
One possible reason for the higher Pass@1 scores is that the models might have more changes to learn from a vast set of samples in this category. 


On the other hand, LLMs struggle to generate code for the \textit{Model Construction} stage with the
lowest Pass@1 scores.
This is because the code for this stage is more complex, very project-specific, and often longer.

\begin{tcolorbox}[boxrule=0.5pt, colback=gray!10, arc=4pt,left=3pt,right=3pt,top=3pt,bottom=3pt,boxsep=0pt
]
\textbf{Finding 2:} Our study shows that LLMs perform best (e.g., GPT-4o's Pass@1 score is 33\%) in Pre/post Processing stages because code in such stages involves common, repetitive tasks, making them easier to learn and generate. In contrast, the Model Construction stage has the lowest scores (e.g., GPT-4o's Pass@1 is 26\%) due to its complexity, variability across projects, and the need to integrate multiple libraries. 
\tool enables detailed analysis which helps future work develop prompting techniques to provide LLMs with the appropriate context and information, improving their performance.
\end{tcolorbox}

\input{tabs/RQ4_ACC}

\textbf{Are different ML task types easier or harder to generate code for?}
In this analysis, we demonstrate how \tool's categorization of ML task types can enable deeper analysis of LLMs code generation and may provide additional insight that can help research propose more accurate prompting techniques and models. Specifically, we made use of our categorization of ML/DL task types: \textit{Classification}, \textit{Regression}, \textit{Object Detection}, \textit{Image Segmentation}, \textit{Time Series Prediction}, \textit{Recommendation}, and \textit{General}.

There is a significant disparity in the Pass@1 score of generated code across task types. Notably, scores for \textit{Recommendation} tasks were the highest among all LLMs, with the best score of 58\% with GPT-4o. On the other end of the scale, \textbf{Image Segmentation} tasks' scores are the lowest across all LLMs. These results indicate that each task type has characteristics LLMs can or not yet capture.

\begin{tcolorbox}[boxrule=0.5pt, colback=gray!10,  arc=4pt,left=3pt,right=3pt,top=3pt,bottom=3pt,boxsep=0pt
]
\textbf{Finding 3:} Different ML/DL tasks vary in complexity affecting LLMs' code generation abilities.
LLMs score highest (up to 58\% with GPT-4o) for recommendation code due to their predictable input structure and patterns.
However, they struggled with image segmentation tasks (up to only 21\%) since they require pixel-level understanding and generalization across variable inputs. \tool's categorization provides insights to improve DL code generation techniques.
\end{tcolorbox}

\input{tabs/RQ3_ACC}

\textbf{Do the various required input data types have any effect on how LLMs generate DL code?}
This analysis aims to investigate if different input data types have different effects on how well LLMs generate code. \tool enables this analysis by providing a categorization of input data types: \textit{Image}, \textit{Text}, \textit{Structured Array}, and \textit{Others}. By comparing their performance across these input types, the study evaluates 
the versatility and limitations of LLMs in dealing with varied data sources.
Table~\ref{tab:rq3_accuracy} shows the Pass@1 of all LLMs 
across different types of input data.
Specifically, performance for image-related tasks is the lowest (only up to 25\% for GPT-4o).
This can be attributed to the inherent complexity and lack of consistent structure in image data, such as varying shapes, resolutions, and channel configurations (e.g., grayscale vs. RGB). 


On the other hand, our results show that textual input data type in code generation exhibits better performance (up to 32\%). We assume that most textual input data type are tokenized and converted before being processed in the DL model, which makes functions that deal directly with textual input data types quite standard and easier to generate.
The structured array category shows slightly higher performance (up to 29\% with GPT-4o) compared to image data.
This is because structured data inherently provides a clearer, more organized format, reducing the reasoning required by the model. As a result, the model more easily generates accurate code for table-related tasks, as opposed to the unstructured, abstract nature of image-based tasks. This suggests that structured data simplifies the generation process.



\begin{tcolorbox}[boxrule=0.5pt, colback=gray!10,  arc=4pt,left=3pt,right=3pt,top=3pt,bottom=3pt,boxsep=0pt
]
\textbf{Finding 4:} This RQ demonstrates the detailed analysis possible with \tool. Our study shows image data had the lowest performance (Pass@1 score up to 25\%) due to complex input structures. In contrast, textual data tasks achieved higher performance (up to 32\%), likely because of more deterministic coding in the pre-processing stages.
\end{tcolorbox}

%% file: tabs/RQ1_ACC.tex
\begin{table}[]
\caption{Pass@1 (\%) scores for various SOTA LLMs on DS-1000 and \tool.}
\label{tab:RQ1_ACC}
\center
\resizebox{0.6\linewidth}{!}{
\begin{tabular}{llr}
\toprule
\textbf{Benchmark}                                      & \textbf{LLM}              & \textbf{Pass@1} \\ 
\midrule
\multirow{4}{*}{DS-1000}& GPT-4o           & 60\\
 & Claude 3.5 sonnet&54\\
 & LLama 3.1 70B&41\\
 & Mistral 7B&20\\
\midrule
\multicolumn{1}{l}{\multirow{4}{*}{\tool}}  & GPT-4o           & 31   \\
\multicolumn{1}{c}{}                         & Claude 3.5 sonnet & 28   \\
\multicolumn{1}{c}{}                         & LLama 3.1 70B    & 21   \\
\multicolumn{1}{c}{}                         & Mistral 7B       & 15   \\
\bottomrule
\end{tabular}
}
\end{table}

%% file: tabs/RQ2_ACC.tex
\begin{table}[]
\caption{Pass@1 scores (\%) for different LLMs in different stages on \tool}
\label{tab:RQ2_MLStage}
\resizebox{\linewidth}{!}{
\begin{tabular}{@{}l@{}rrrrr@{}}
\toprule
\textbf{Model}                        & \makecell{ \textbf{Pre/Post} \\ \textbf{Processing} }    & \makecell{ \textbf{Model} \\ \textbf{Construction}} & \textbf{Training}              & \textbf{Inference}            & \makecell{ \textbf{Evaluation} \\ \textbf{\& Metrics} }  \\ \midrule
GPT-4o           & \textbf{33} & \textbf{26} & \textbf{31} & 30&\textbf{32} \\
Claude 3.5 sonnet & 31 & 22 & 28 & 30& 29 \\
LLama 3.1 70B    & 22 & 14 & 21 & 29 & 24 \\
Mistral 7B          & 14 & 11& 16& 26 & 19 \\
\bottomrule
\end{tabular}
}
\end{table}

%% file: tabs/RQ4_ACC.tex
\begin{table*}[h]
    \centering
    \caption{Pass@1 (\%) scores for different ML/DL tasks on \tool}
    \label{tab:rq4_accuracy}
    \resizebox{\linewidth}{!}{
    \begin{tabular}{@{}lrrrrrrr@{}}
    \toprule
    \textbf{Model} & \textbf{Classification} & \textbf{Regression} & \textbf{Object Detection} & \textbf{Image Segmentation} & \textbf{Time Series Prediction} & \textbf{Recommendation} & \textbf{General} \\ 
    \midrule
    GPT-4o &
      30&
      \textbf{36}&
      28&
      19&
      \textbf{33}&
      \textbf{58}&
      \textbf{31} \\
    Claude 3.5 sonnet&
      \textbf{32} &
      35&
      \textbf{30} &
      \textbf{21}&
      29&
      47&
      27\\
    LLama 3.1 70B &
      28 &
      22&
      17 &
      14 &
      31&
      40&
      19 \\
    Mistral 7B &
      24 &
      21 &
      11 &
      11 &
      13 &
      29&
      13 \\
      \bottomrule
    \end{tabular}
    }
\end{table*}

%% file: tabs/RQ3_ACC.tex
\begin{table}[h]
    \centering
    \caption{Pass@1 (\%) scores across various input data types on \tool}
    \label{tab:rq3_accuracy}
    \resizebox{\linewidth}{!}{
    \begin{tabular}{@{}lrrrr@{}}
    \toprule
    \textbf{Model}                        & \textbf{Image}                 & \textbf{Text}                  & \textbf{Structured Array}               & \textbf{Others} \\ \midrule
    GPT-4o          & \textbf{25}& \textbf{32}& \textbf{29} &\textbf{ 40}\\
    Claude 3.5 sonnet& \textbf{25}& 30& 24 & 33\\
    LLama 3.1 70B    & 18 & 30 & 19& 26 \\
    Mistral 7B          & 8& 23 & 13 & 25\\
    \bottomrule
    \end{tabular}
    }
\end{table}

%% file: 6_qualitative.tex
\section{Qualitative Analysis}

\input{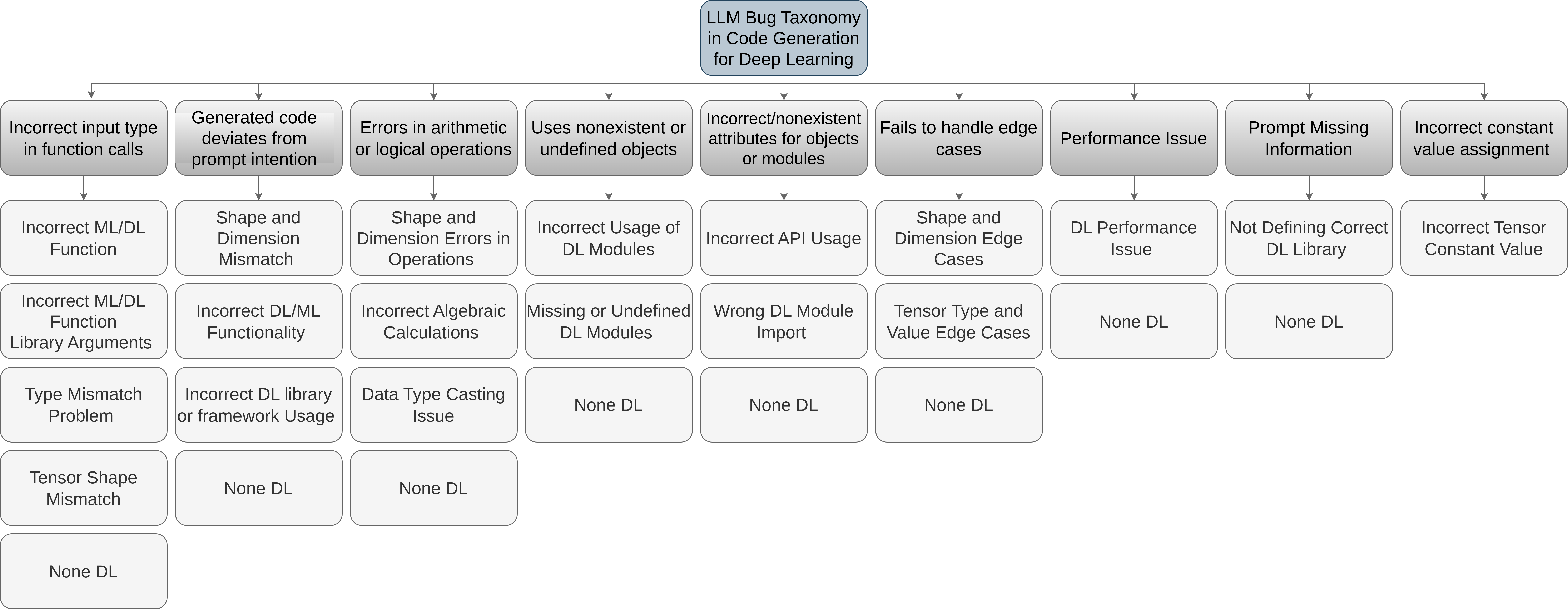}
\input{figs/tax_distribution}

\textbf{Taxonomy of Bugs in Generated DL Code:} GPT-4o achieves a pass@1 rate of only 31\% hence in most cases, it failed to generate the correct code. In this section, we build a taxonomy of common bug patterns and issues that arise in DL code generated by GPT-4o. This taxonomy is an expansion of Tambon et.al \cite{tambon2024bugs}'s bug taxonomy for LLM-generated regular code.

Following the same procedure as our labeling process, three authors manually investigate all GPT-4o failures and categorize them following Tambon et.al's taxonomy. At the same time, the annotators identify the DL-specific sub-categories for each failure. The result is the taxonomy presented in Fig~\ref{fig:taxonomy}. The appendix gives a detailed explanation of each bug type and sub-category.



\textbf{Failures in Generating the DL Code and General Code:} Tambon et. al\cite{tambon2024bugs} analyzed failures when CodeGen models generate code for the general tasks. Figures~\ref{fig:tax_distribution} show the distributions of the bug types when generating general code vs DL code. 
On the one hand, \textit{misinterpretation} (purple) is a common bug when generating both general code and DL code, however, due to more complex logic and arithmetic requirements, LLMs more often make this mistake when generating DL code. An example of this type of bug can be seen in Figure~\ref{fig:discussion2}. On the other hand, since GPT4o is much more capable compared to CodeGen models used by prior work, errors such as \textit{incomplete generation} (green), \textit{silly mistake} (dark gray), and \textit{syntax error} (yellow) occur at a much lower rate.

Furthermore, we have introduced several new categories of bugs that commonly arise in DL code generation. Firstly, \textit{errors in arithmetic and logical operations}(light blue) occur when incorrect calculations or flawed logical code are generated.
Secondly, \textit{performance}(light brown) issues involve inefficient generated code with slow execution times, excessive memory consumption, or suboptimal utilization of resources.
Lastly, \textit{prompt missing information}(light purple) when the prompts are missing details to fully address the problem at hand, resulting in incomplete or partially implemented solutions.
These new categories identify important challenges that are unique to DL code generation.

\begin{tcolorbox}[boxrule=0.5pt, colback=gray!10,  arc=4pt,left=3pt,right=3pt,top=3pt,bottom=3pt,boxsep=0pt
]
\textbf{Observation 1:} \textit{Misinterpretation} is a common issue in both generated general code and DL code, however, due to more complex logic and arithmetic requirements, LLMs are more likely to make this mistake when generating DL code. \textit{Errors in arithmetic and logical operations}, \textit{performance}, and \textit{prompt missing information} emerged as new issues that are specific to DL code generation.
\end{tcolorbox}

\input{figs/discussion3}
\newpage
\textbf{Bugs in Human-Written vs. LLM-Generated DL Code:} On one hand, prior study~\cite{islam2019comprehensive} identified the most common types of bugs in human-written DL code which include logic errors, API misuse, and data-related issues. Among these, API misuse is the most prevalent bug pattern in human-written DL code when using TensorFlow, whereas data flow bugs are more common when using PyTorch. On the other hand, according to our analysis of LLM-generated DL code, although API misuse remains a frequent issue, data structural problems, such as tensor mismatches and dimensional errors, occupy more frequently. Figure~\ref{fig:discussion3} highlights an instance of dimensional mismatches in LLM-generated DL code. In this case, GPT-4o incorrectly assumes that each shift value can be applied directly to all pixels in the image channel, causing a shape mismatch.


\input{figs/discussion2}
Similar to prior findings~\cite{islam2019comprehensive} of human-written DL code, LLM-generated DL code often contains  
logic errors.
This similarity may stem from the fact that LLMs are trained on human-written code, thereby inheriting logical structures and concepts from human programmers. An example of such logic-related bugs is shown in Figure~\ref{fig:discussion2}, demonstrating how LLMs replicate logical reasoning errors that occur in human-written code. Here, GPT-4o applies \textit{scale\_x} only to the cosine whereas the scaling factors \textit{scale\_x} and \textit{scale\_y} should be applied uniformly to both the sine and cosine components of the rotation matrix. This results in improper scaling along the axes and triggers a test failure.

\input{figs/discussion1}
Additionally, API misuse is a common bug pattern occurred in both human-written and LLM-generated DL code.
Figure~\ref{fig:discussion1} provides an example of API misuse in LLM-generated code where GPT-4o attempts to call \textit{torch.idct}, which is not implemented in PyTorch. One possible fix is to provide more context concerning third-party libraries. For example, one could hint to LLMs to use \textit{scipy} instead, resulting in \textit{scipy.fftpack.idct(x.numpy(), norm=norm)} instead.

\begin{tcolorbox}[boxrule=0.5pt, colback=gray!10,  arc=4pt,left=3pt,right=3pt,top=3pt,bottom=3pt,boxsep=0pt
]
\textbf{Observation 2:} Unlike human-written DL code, LLM-generated DL code contains more data structural problems, such as tensor mismatches and dimensional errors. 
Similar to prior findings of human-written DL code, LLM-generated DL code often contains logic errors.
Additionally, API misuse frequently occurs as a bug pattern in both human-written and LLM-generated DL code.
These overlaps suggest that while LLMs exhibit unique weaknesses, their reliance on human-generated training data also leads to shared bug patterns, particularly in logic and API misuse errors.

\end{tcolorbox}

%% file: figs/taxonomy_graph.tex
\begin{figure*}[t!]
    \centering
    \includegraphics[width=\linewidth]{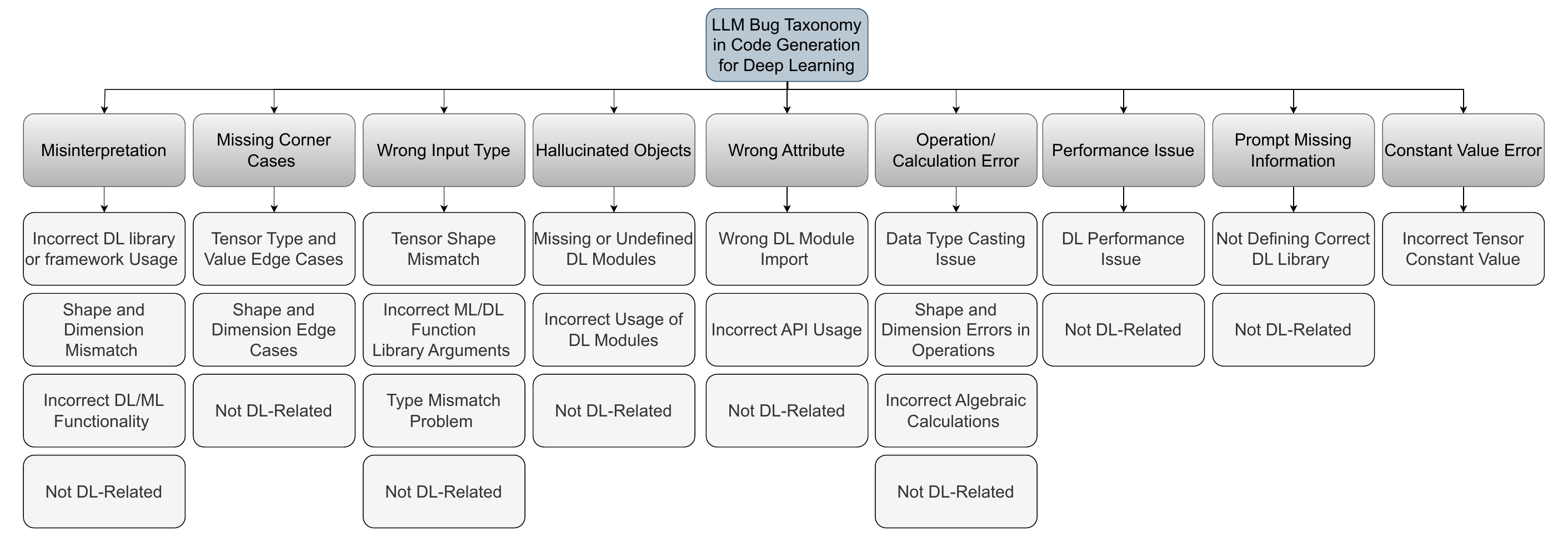}
    \caption{\centering Taxonomy of bugs in DL generated code. Only categories with DL-related subcategories are shown.
    }
    \label{fig:taxonomy}
\end{figure*}

%% file: figs/tax_distribution.tex
\begin{figure*}[t!]
    \centering
    \includegraphics[width=\linewidth]{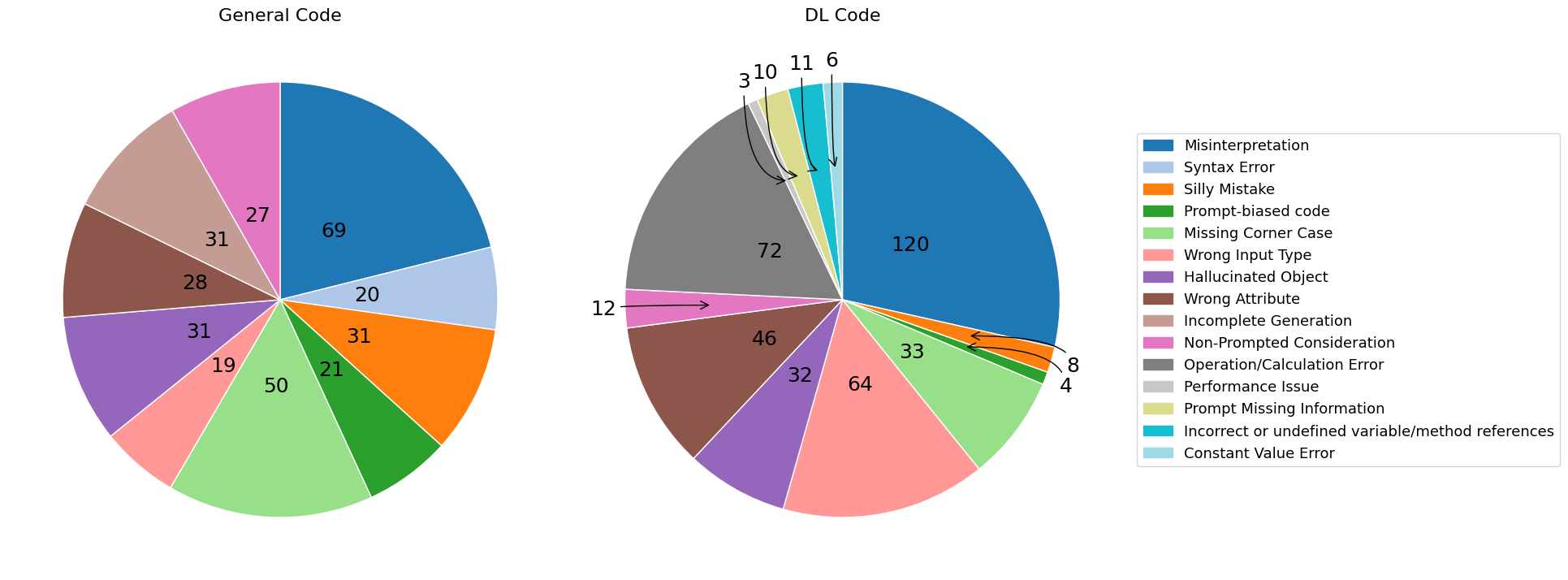}
    \caption{Distribution of bugs in general code vs. DL code generated by LLM
    }

    \label{fig:tax_distribution}
\end{figure*}

%% file: figs/discussion3.tex
\begin{figure}
    \centering
    \includegraphics[width=1\linewidth]{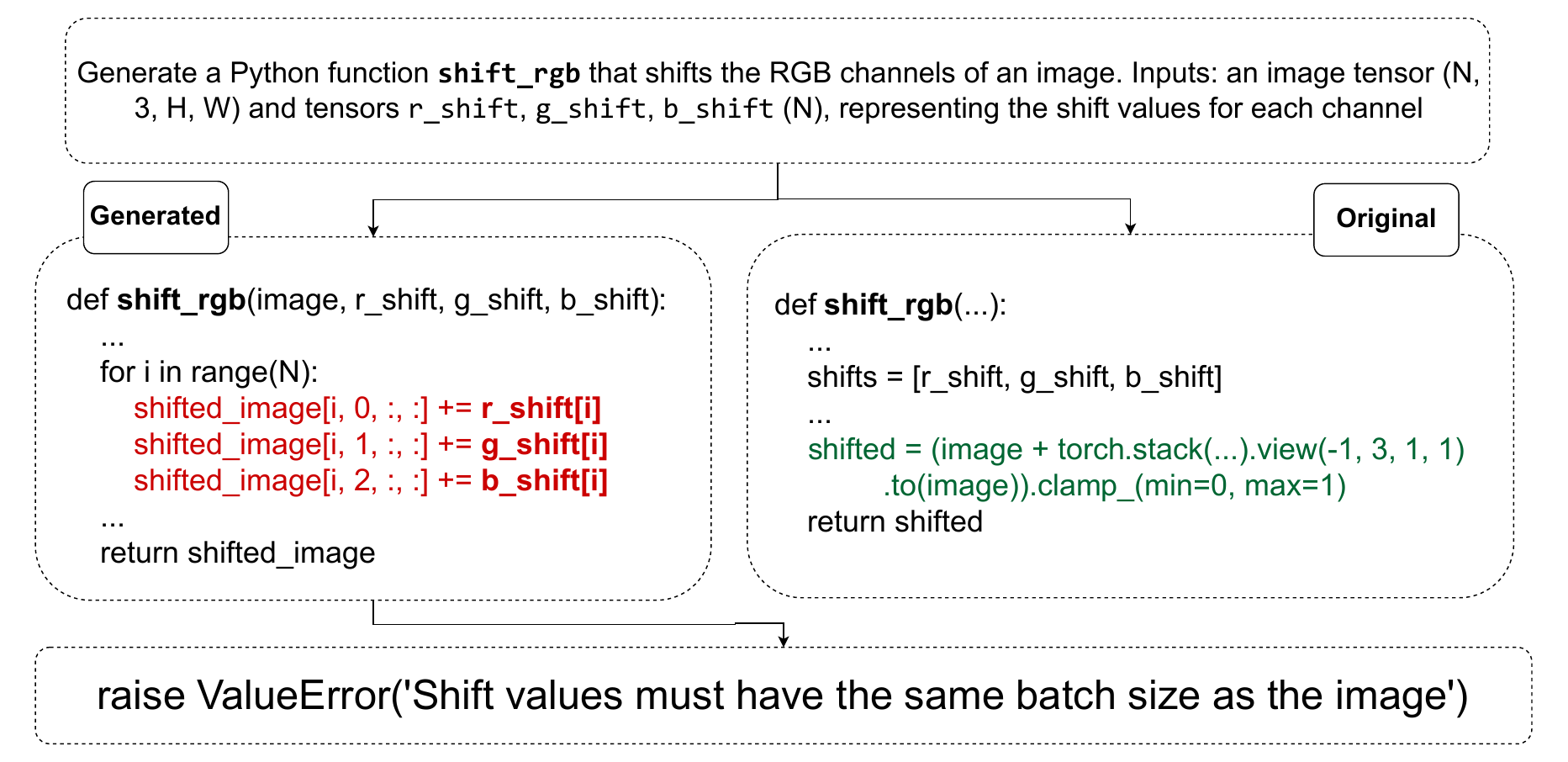}
    \vspace{-7pt}
    \caption{Mismatching data shapes: shifting variables need to be broadcasted to the image shape
    }
    \label{fig:discussion3}
    \vspace{-10pt}
\end{figure}

%% file: figs/discussion2.tex
\begin{figure}
    \centering
    \includegraphics[width=1\linewidth]{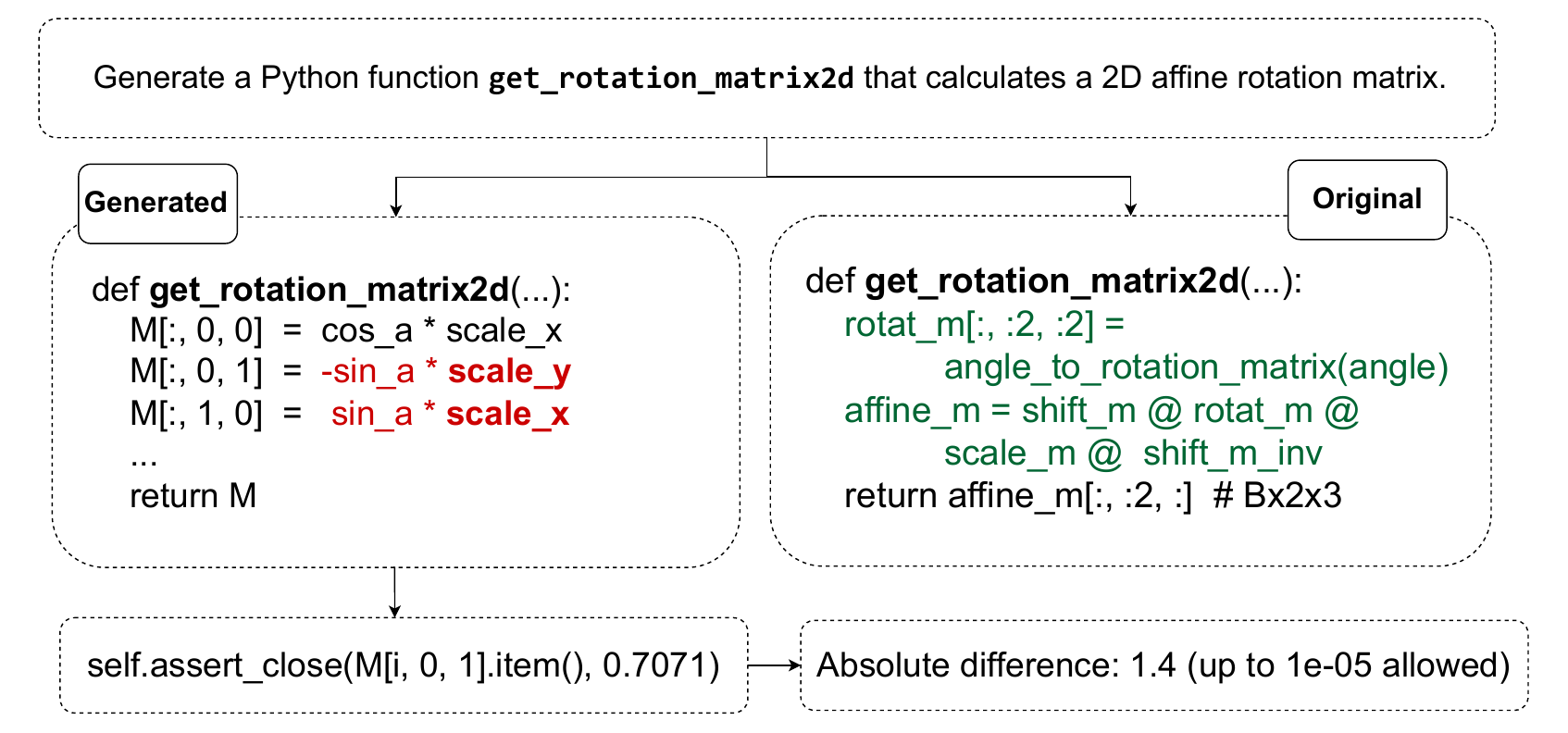}
    \vspace{-7pt}
    \caption{Incorrect processing of parameters: The axes scales need to be applied to both sin and cos
    }
    \label{fig:discussion2}
    \vspace{-10pt}
\end{figure}

%% file: figs/discussion1.tex
\begin{figure}
    \centering
    \includegraphics[width=1\linewidth]{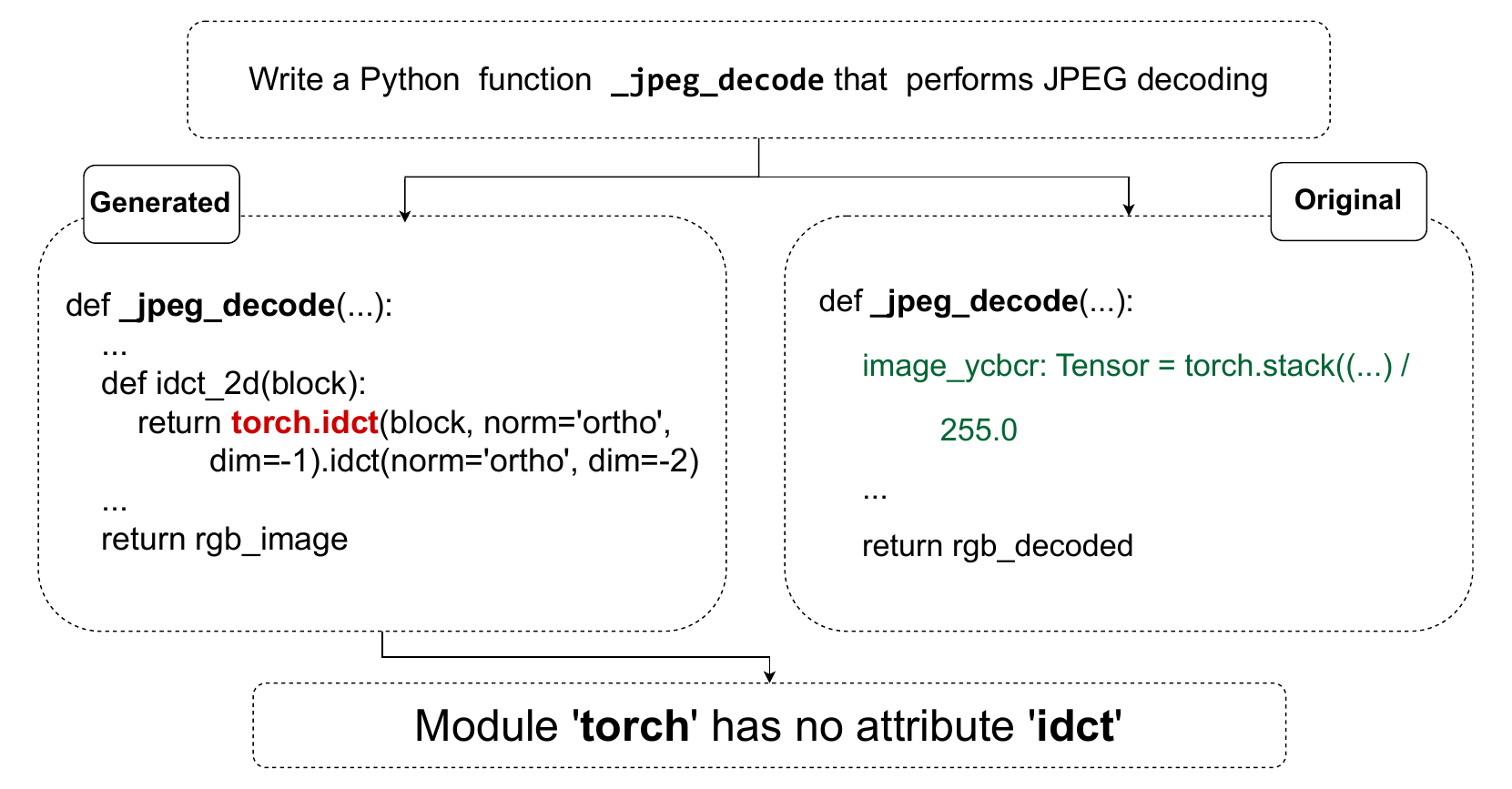}
    \caption{Wrong usage of a third-party library. 
    }

    \label{fig:discussion1}
    \vspace{-10pt}
\end{figure}

%% file: 7_limit.tex
\section{Threats to validity}

Even with the temperature parameter set to zero, our experiments still utilized non-deterministic models. While a lower temperature reduces randomness, it does not fully eliminate variability in the models' outputs~\cite{ouyang2023llm,song2024good}. Also, we sourced data from various repositories related to DL and AI but did not include all possible repositories or tags. Expanding the dataset could capture a wider range of use cases and code patterns. 

Data labeling was performed by four independent annotators, achieving strong inter-rater reliability. Despite this, some labeling conflicts persisted and were addressed through discussions to reach a consensus. However, not all discrepancies could be fully resolved. Also, even if we used the commonly used pass@k metric to evaluate model performance, prior research~\cite{shiri2024history} shows that passing all test cases does not guarantee complete code correctness, especially in edge cases.

%% file: 8_conclusion.tex
\section{Conclusion}
In this paper, we introduce \tool, a benchmark for deep learning tasks related to code generation. The dataset comprises 520 instances, gathered from the most starred and recently updated GitHub repositories. We categorize the data based on the pipeline stage, ML task, and input data type. 
Additionally, our quantitative analysis of the performance of four state-of-the-art LLMs on \tool reveals that DL code generation is challenging and \tool can provide more insight to help improve the generation process.
Using our taxonomy of issues found in LLM-generated DL code, the qualitative analysis reveals the distinct challenges that LLMs face when generating DL code compared to general code as well as the similarities and differences between human-written and LLM-generated DL code.



%% file: 10_appendix.tex
\appendix
\onecolumn

\noindent {\Large  \textbf{Appendix}}

Table of Contents: 
\begin{itemize}
    \item Appendix A: Dataset Statistics
    \item Appendix B: Candidate Prompt Filtering Criteria
    \item Appendix C: Final Data Filtering and Validation Criteria
    \item Appendix D: Data Categories
    \item Appendix E: LLM Bug Types and DL-specific Subtypes
    \item Appendix F: Distribution of Failures in Generated DL Code
    
\end{itemize}

\section{Dataset Statistics}

\input{figs/statistics_pie}

\tool consists of 520 instances of AI and DL data points (filtered from over 2,000 raw data points). The data is curated from 30 GitHub repositories (selected from an initial pool of 160 related repositories).
To ensure an accurate evaluation of code generation techniques under test,
each prompt instance in \tool is accompanied by
at least three test cases (six test cases on average).
One of \tool's contributions is the categories that we assign for each data point. As mentioned in Section~\ref{sec:BenchConstr}, each data point is assigned a label for which stage of the ML pipeline it belongs to, a label for which ML task it helps solve, and a label for the type of input data. This information enables users of our benchmark to perform an in-depth analysis of their proposed technique with respect to multiple ML-specific aspects. We demonstrate this in our empirical study presented in Section~\ref{sec:study} later.

\input{figs/processing}
\input{figs/model_construction}

Fig~\ref{fig:statistics} represents the distribution of \tool's data in each categorization. In terms of the stages in the ML pipeline (Fig~\subref{fig:statistics_step}), our dataset well covers the five stages of the ML pipeline with the pre/post-processing stage having the most (210) representative samples. Fig~\ref{fig:preproprocessing} lists the prompt to generate 
a pre/post-processing ``draw\_point2d'' function that can be used to 
highlight key points of interest in output images. The model construction stage contains the second-most (119) samples such as the one shown in Fig~\ref{fig:model_construction}. This example shows the prompt to generate the
``\_\_init\_\_'' method for a fully connected neural network (FCNN). Other ML stages have an equal share of samples. This indicates a balanced dataset that covers all ML stages.


\input{figs/general_example}
\input{figs/example_classification}

Most of our data serve more than one ML task type, hence 328 (over 63\%) instances are labeled as \textit{General} as shown in Fig~\subref{fig:statistics_task}. For example, Fig~\ref{fig:example_general} shows \textbf{to\_image} function handles data type conversions and pre-processing to standardize image inputs, without performing any specific machine learning task. However, for the cases that serve a specific ML task, our dataset covers all ML tasks evenly with 14 to 60 instances each. Among these, the classification task has the most representative of 60 data points. For example, Fig~\ref{fig:example_classification} shows a classification task, calculating precision, recall, and F1 scores for both duplicate and non-duplicate file pairs to evaluate the performance of a classification model. On the other hand, The regression task is not as popular with only 14 data points.


Image data is the most popular input data type with 238 instances (nearly 46\%) as shown in Fig~\subref{fig:statistics_data}. In some cases where the input data to the function is missing or not the input to the model, we categorize them into the Others category which contains 168 instances. An example of such cases is presented in Fig~\ref{fig:model_construction}, where the initialization method constructs a new neural network model, however, information on the input type of such networks is not available. Textual data has the least instances since most of the time, textual data is tokenized and presented as either a data array or general tensor.


\section{Candidate Prompt Filtering Criteria}
In this appendix, we describe the criteria of filtering and refining prompts to ensure clarity and completeness.

\begin{description}
\item[Contains clear sufficient information for the code to be generated]
This assessment aims to ensure the prompt’s clarity and comprehensibility for a human expert. Annotators check that the prompt includes all essential variables, functions, and definitions for high-quality code generation, providing enough information to clearly explain the problem. The human expert serves as the benchmark to set a high standard for future code generation. We also verify that the prompt provides sufficient guidance, including specific coding conventions or required components.


\item[Specifies the input and output format]
Since our test cases require certain input and output formats, it is important to check such details in the candidate prompt to enable our test cases to function correctly\cite{sahoo2024systematic, chen2024using}. In other words, without precise definitions of the input and output specifications, the generated code might not align with the expected test parameters, resulting in false negative results during evaluation. Error and exception handling are also considered in this question. For example, we specifically check whether the prompt accounts for handling cases such as ``ValueError'', ``TypeError'', or other domain-specific exceptions that the function might raise. This will ensure that the code will be correctly evaluated given our extracted test cases.


\item[Covers error handling and boundary conditions]
Similar to input and output specification, error handling and boundary conditions are often part of the required testing parameters
By ensuring that the prompt includes such details, we ensure that the passing rate truly reflects the performance of the code generation under test.
\end{description}

\section{Final Data Filtering and Validation Criteria}

This appendix outlines the criteria used to filter and validate data, ensuring alignment with key DL tasks, proper use of AI frameworks, and clarity in algorithm implementation.

\begin{description}
\item[\textbf{Serving key DL tasks}] The prompt and the associated function should be closely aligned with significant DL tasks such as image recognition, regression, item recommendation, object detection, label prediction, and natural language processing tasks. This criterion ensures that our dataset contains all important and relevant data points\cite{xie2024frontiers}.

\item[\textbf{Utilization of popular DL frameworks}] The code should efficiently use widely recognized AI frameworks (when appropriate), such as TensorFlow, PyTorch, or Keras. This criterion ensures our dataset represents typical DL code with a heavy emphasis on reusability\cite{assi2024unraveling}.

\item[\textbf{Algorithms' relevancy and clarity}] The code should implement DL-specific algorithms (e.g., edge detection algorithms, Principal component analysis, or Stochastic gradient descent). The code should also be well-documented and easy to understand. Complex algorithms must strike a balance between technical depth and clarity to ensure usability.
\end{description}

\section{Data Categories}
In this appendix, we provide details of three key sample categorizations: the stage in the ML pipeline, the ML task type, and the input data type. 

\subsection{Stage in the ML pipeline}

    This label indicates the stage that the code is in within the ML pipeline: \textit{Pre/post Processing}, \textit{Model Construction}, \textit{Training}, \textit{Inference}, or \textit{Evaluation \& Metrics}.
    The annotators determine whether the function is related to a stage by analyzing the code and comment to find information that is related to the specific stage. For example, code that specifies a convolutional neural network (CNN) architecture with layers such as convolutions or pooling would fall under the Model Construction category. 

\begin{description}
    \item[\textbf{Pre/Post Processing}] Code in the pre or post-processing stage often manipulates data (input or output). For example, pre-processing code cleans or augments input data, whereas post-processing code augments output data for visualization. Due to the ambiguity at the function level, we have a combined category for pre and post-processing code\cite{wen2020time}.

    \item[\textbf{Model Construction}]This stage defines the network architecture and sets up the computational graph for deep learning models, including defining layers, activation functions, and layer connections. Examples include defining CNN architectures and forward pass logic. Loss functions are part of this stage, but optimization steps are in the training phase\cite{howard2019searching}.

    \item[\textbf{Training}]The training stage optimizes the model's parameters using a loss function and optimization algorithm. This includes backpropagation and weight updates. Code for gradient descent using optimizers like Adam or SGD and looping over epochs and batches falls under this stage\cite{diederik2014adam}.

    \item[\textbf{Inference}]Inference code is used to generate labels based on a trained model. It processes new input data and outputs results, such as classifications or detections, without changing model parameters. This stage emphasizes speed and efficiency for real-time deployment\cite{kirillov2019panoptic}.

    \item[\textbf{Evaluation \& Metrics}]Code in this stage assesses the performance of a trained model using various metrics. It involves running the model on a validation/test dataset and comparing predictions to ground truth labels to measure accuracy, precision, recall, F1-score, etc.\cite{wu2020comprehensive}.

\end{description}

\subsection{ML task type}
    
This label indicates the ML task\cite{sarker2021machine, Vinodkumar2023Survey, technologies2024Manakitsa} that the code is serving when applicable. The annotators examine the code to determine the type of task being solved, such as \textit{Time series Prediction}, \textit{Recommendation}, \textit{Image Segmentation}, \textit{Object Detection}, \textit{Regression}, \textit{Classification}, or \textit{General}. Specifically, the annotators look for patterns in the code corresponding to each task. For instance, code that outputs bounding boxes and class labels for objects falls under the Object Detection category. In cases where the code can be used for multiple ML tasks (i.e., does not exclusively belong to a specific ML task), we assigned a \textit{General} label.

\begin{description}

    \item [\textbf{Classification}]Classification tasks involve assigning input data to categories or classes. For example, models using softmax activation in the final layer for outputs like ``dog'' or ``cat'' fall under this category. Categorical cross-entropy loss is a common indicator.

    \item [\textbf{Regression}]Regression tasks predict continuous values. Code indicating regression tasks often has linear activation functions in the final layer.

    \item [\textbf{Object Detection}]Detection tasks identify objects and their locations within images. Code that outputs bounding boxes and class labels (e.g., YOLO, Faster R-CNN) and employs anchor boxes or non-maximum suppression is indicative of detection tasks.

    \item [\textbf{Image Segmentation}]Segmentation tasks assign labels to each pixel in an image. Code involving semantic or instance segmentation (e.g., U-Net, Mask R-CNN) where the output is a mask with pixel-level classifications is a common example.

    \item [\textbf{Time Series Prediction}]These tasks forecast future values using historical data. Code involving recurrent neural networks (RNNs), LSTM, GRU models, and loss functions like mean absolute error (MAE) or MSE is typical.

    \item [\textbf{Recommendation}]Recommendation tasks suggest items or actions based on user data. Code implementing collaborative or content-based filtering algorithms, matrix factorization, or deep learning-based models for recommendations falls into this category.

    \item [\textbf{General}]Code that is versatile and applicable to multiple ML tasks without being exclusive to a specific one is labeled as \textbf{General}.

\end{description}

\subsection{Input data type}

This label indicates the input data type of the function. We focus on typical ML input data types such as \textit{Image}, \textit{Text}, \textit{Structured Array} (i.e., tabular), and \textit{Others}. The annotators analyze the processing flow of data to assign accurate labels. For example, techniques like flipping, cropping, or adding noise process image input. When the input data does not fit one of the typical types (image, text, structured array), we assign the Others label. 

\begin{itemize}
    
    \item \textbf{Image}---Processing for image data includes steps like resizing, normalization, and data augmentation. Code that resizes images (e.g., 224$\times$224 for CNNs), normalizes pixel values, or applies augmentations (flipping, cropping, noise addition) typically signals image data\cite{krizhevsky2012imagenet}.

    \item \textbf{Text}---Text processing involves tokenization, n-gram generation, stemming, lemmatization, and embeddings. Code that handles these processes and converts text into vectors (e.g., using TF-IDF, Word2Vec, BERT) indicates text data\cite{liu2018neural}.

    \item \textbf{Structured Array}---Tabular data, where rows represent data points and columns represent features, is processed by normalization, one-hot encoding, or handling missing values. Code that reads CSVs into DataFrames and applies these techniques indicates structured array data, commonly used in regression or classification tasks\cite{chen2016xgboost}.

    \item \textbf{Others}---When input data does not match typical types (image, text, structured array), it is labeled as \textbf{Others}. This includes input such as model parameters or hyperparameters. For example, \verb|def __init__(self, weight, bias=None)| initializing model components without direct input data processing falls under this label.

\end{itemize}

\section{LLM Bug Types and DL-specific Subtypes}

In this appendix, we provide details for the common types of errors in LLM-generated code as well as our DL-specific subtypes.

\begin{description}

\item[Misinterpretation: \textit{Generated code deviates from the prompt intention}]  
The produced solution does not fulfill the user’s original requirements or strays from the specified goals. This often indicates that the LLM has misunderstood or incompletely parsed the prompt.
    
\begin{description}
\item[Incorrect DL Library or Framework Usage:] The generated code does not match the requested library or framework. For example, if the prompt asks for a TensorFlow implementation of a CNN, but the LLM generates the model using PyTorch instead, or if a user requests a NumPy-based neural network operation but the output code uses TensorFlow functions.

    \item[Shape and Dimension Mismatch:] The LLM produces code with incorrect tensor dimensions that do not follow the prompt specifications. For example, if the prompt requests a fully connected layer expecting an input of shape $(64, 128)$, but the generated code initializes it with an input shape of $(128, 64)$, leading to a mismatch in matrix operations.

    \item[Incorrect DL/ML Functionality:] The generated code does not implement the correct functionality as described in the prompt. For instance, if the prompt asks for a binary classification model using a sigmoid activation function, but the output code instead applies a softmax activation function intended for multi-class classification, altering the intended behavior.

\end{description}
    
\item[Syntax Error: \textit{Missing parenthesis, semicolon, or other syntax issues}]
Straightforward syntactic mistakes such as unclosed quotes, unmatched braces, or misplaced punctuation prevent the code from compiling or running properly.

\item[Silly Mistake: \textit{Redundant conditions, unnecessary casting}] 
Simple but avoidable errors, such as repeating the same condition twice or performing extra type conversions with no purpose. While these do not always break the code, they reduce readability and hint at confusion in the model's reasoning.

\item[Prompt-biased Code: \textit{Code overly relies on examples from the prompt}]
The LLM anchors too strongly to the examples provided in the prompt, resulting in a solution that works only for the specific inputs shown rather than generalizing the logic for broader applicability.

\item[Missing Corner Cases: \textit{Edge cases not handled}]
The generated solution neglects special scenarios such as empty inputs, boundary values, or invalid parameters, leading to unreliable behavior outside of typical inputs.

\begin{description}

    \item[Tensor Type and Value Edge Cases:] These bugs occur when operations fail due to unexpected tensor types or values. For example, using a tensor with \texttt{float32} data type in a function that expects integers or encountering issues when dividing by zero in a tensor.
    
    \item[Shape and Dimension Edge Cases:] Bugs of this type happen when operations fail because of unexpected edge-case shapes. For example, trying to perform a convolution on a tensor with a batch size of $0$ or a single dimension, such as $(1, 28, 28)$, when a shape like $(32, 28, 28)$ is expected.
\end{description}

\item[Wrong Input Type: \textit{Incorrect input type in function calls}]
The code passes incompatible data types to functions or methods (e.g., providing a string instead of a list), which causes runtime failures or nonsensical outputs.

\begin{description}
    \item[Tensor Shape Mismatch:] The generated code provides tensors with incorrect shapes to functions, leading to shape-related errors. For example, passing a 3D tensor of shape $(batch, height, width)$ to a function that expects a 4D tensor of shape $(batch, channels, height, width)$, causing a runtime error in deep learning frameworks like PyTorch or TensorFlow.
    
    \item[Incorrect ML/DL Function Library Arguments:] These occur when invalid arguments are passed to functions. For instance, using \texttt{stride=-1} in a convolution function, which is not logically or mathematically valid.

    \item[Type Mismatch Problem:] The generated code uses tensors with incompatible data types in operations. For example, passing a tensor with data type \texttt{float32} to a function that expects \texttt{int64}, or attempting to index a tensor with a floating-point value instead of an integer, leading to type-related execution failures.
\end{description}

\item[Hallucinated Object: \textit{Nonexistent or undefined objects used}] 
The LLM invents objects, classes, or modules that do not exist or have not been imported or defined. These errors result in runtime failures or developer confusion.
\begin{description}
    \item[Missing or Undefined DL Modules:] This happens when a model, layer, or module that hasn’t been properly defined or initialized is used. For example, attempting to forward-pass input through a neural network layer that hasn't been added to the model.
    
    \item[Incorrect Usage of DL Modules:]The generated code references deep learning modules, functions, or classes that do not exist or belong to the wrong framework. For example, calling \texttt{torch.nn.Dense()} instead of \texttt{torch.nn.Linear()}, or attempting to use \texttt{tensorflow.layers.Conv2D} instead of \texttt{tf.keras.layers.Conv2D}. These hallucinated module names cause import errors or incorrect function calls.
\end{description}

\item[Wrong Attribute: \textit{Incorrect/nonexistent attributes for objects or modules}]
The LLM references valid objects but assigns them invalid or incorrect attributes. These subtle errors often result from misunderstandings of library APIs or typos in the generated code.

\begin{description}
 \item[Wrong DL Module Import:] Bugs of this nature arise when modules are imported incorrectly. For example, importing \texttt{jax} functions when the rest of the code is written in PyTorch, leading to incompatibilities during execution.
    \item[Incorrect API Usage:] These bugs occur when a library API function is called incorrectly. For example, using the \texttt{train()} method instead of \texttt{fit()} for a Keras model or passing parameters in the wrong order to an optimizer.
\end{description}

\item[Non-Prompted Consideration: \textit{Non-requested features added}]
The LLM includes functionality unrelated to the requirements, often due to extraneous training data or contextual noise. This bloats the code and complicates its scope.

\item[Operation/Calculation Error: \textit{Errors in arithmetic or logical operations}]
The LLM makes errors in mathematical calculations or logical expressions, such as confusing addition with subtraction or mixing up operator precedence. These subtle mistakes produce incorrect results.

\begin{description}

    \item[Data Type Casting Issues:] These bugs occur when tensors or variables are cast into incompatible data types. For instance, casting a \texttt{float32} tensor into \texttt{int32} without considering the loss of precision, which may disrupt training.
    \item[Shape and Dimension Error in Operations:] The generated code performs mathematical operations on tensors with incompatible shapes or dimensions, leading to incorrect computations or runtime failures. For example, attempting to add two tensors of shapes $(32, 64)$ and $(64, 32)$ without proper broadcasting, or performing a matrix multiplication between tensors with mismatched inner dimensions, such as $(4, 3) \times (5, 4)$, causing a shape misalignment error.
    
    \item[Incorrect Algebraic Calculation:] These bugs refer to mathematical errors in computations. For instance, incorrectly normalizing data by dividing by the mean instead of the standard deviation, leading to improper scaling of input features.

\end{description}

\item[Performance Issue:]
This category includes inefficiencies in the generated code that impact runtime or resource usage. Examples include unnecessary nested loops, unoptimized algorithms, or excessive use of memory. While the code may produce correct results, its suboptimal implementation can make it impractical for large datasets or real-time applications. Performance issues often arise because the LLM generates a brute-force solution without understanding optimization principles.

\begin{description}
    \item[DL Performance Issues:] These bugs refer to inefficiencies in implementation that degrade model performance. For instance, not using GPU acceleration for operations or improper batching strategies leads to high memory consumption and slow training.
\end{description}

\item[Prompt Missing Information: \textit{Incomplete or unclear prompts}]
The bug arises due to insufficient detail or ambiguity in the input prompt, leading the LLM to make assumptions or guess certain details when generating the code. For example, if the prompt does not specify edge case handling or input constraints, the model may overlook these aspects entirely. This highlights the importance of crafting precise and comprehensive prompts when using LLMs for code generation.

\begin{description}
    \item[Not Defining the Correct DL Library in the Prompt:] This occurs when the prompt or instructions fail to specify the appropriate library or framework. For example, a user asks a language model to generate PyTorch code but does not explicitly state this, leading to TensorFlow code generation instead.
\end{description}

\item[Incorrect or Undefined Variable/Method References: \textit{Variables or methods that are not defined or incorrectly referenced}]
The LLM generates code that includes references to variables or methods that do not exist or are improperly used, leading to runtime errors such as NameError or AttributeError.

\item[Constant Value Error: \textit{Incorrect constant value assignment}]
The LLM assigns incorrect or miscalculated constant values, such as setting a time-out period to \texttt{10ms} instead of \texttt{1000ms}, leading to unexpected behavior.
\begin{description}
    \item[Incorrect Tensor Constant Value:] This type of bug arises when tensors are initialized with incorrect values, leading to flawed model behavior. For example, initializing weights or biases with all zeros instead of random values causes issues in training dynamics.
\end{description}

\end{description}

\newpage
\section{Distribution of Failures in Generated DL Code}

Table~\ref{tab:taxonomy} presents the distribution of bugs in LLM-generated DL code. The most prevalent issue is \textbf{deviation from the prompt}, accounting for the largest portion of errors. Unlike general LLM-generated code, DL code is more prone to \textbf{arithmetic and logical errors}, reflecting the complexity of numerical computations. Additionally, \textbf{incorrect input types in function calls} represent a significant share of the identified bugs, highlighting a common source of failures in generated DL code.

\input{tabs/Taxonomies}

%% file: figs/statistics_pie.tex
\begin{figure*}[h!]
  \centering
  \subfloat[Pipeline stages\label{fig:statistics_step}]{
    \includegraphics[width=0.32\linewidth, trim=0 0 0 1cm, clip]{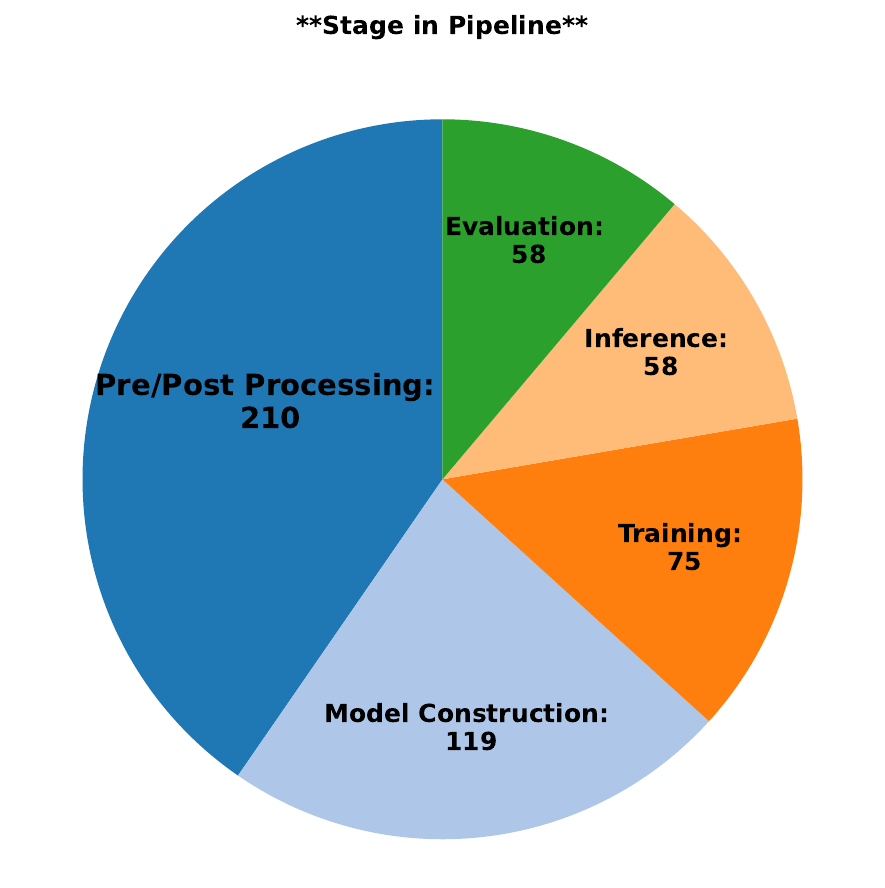}
  }
  \hfill
  \subfloat[Task types\label{fig:statistics_task}]{
    \includegraphics[width=0.32\linewidth, trim=0 0 0 1cm, clip]{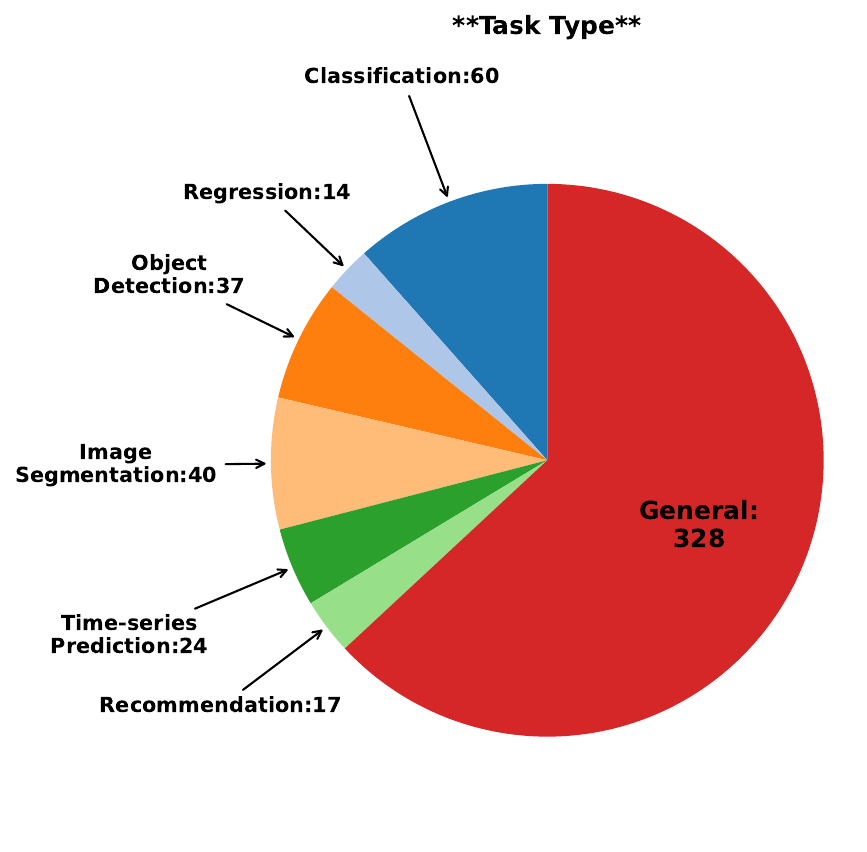}
  }
  \hfill
  \subfloat[Data types\label{fig:statistics_data}]{
    \includegraphics[width=0.32\linewidth, trim=0 0 0 1cm, clip]{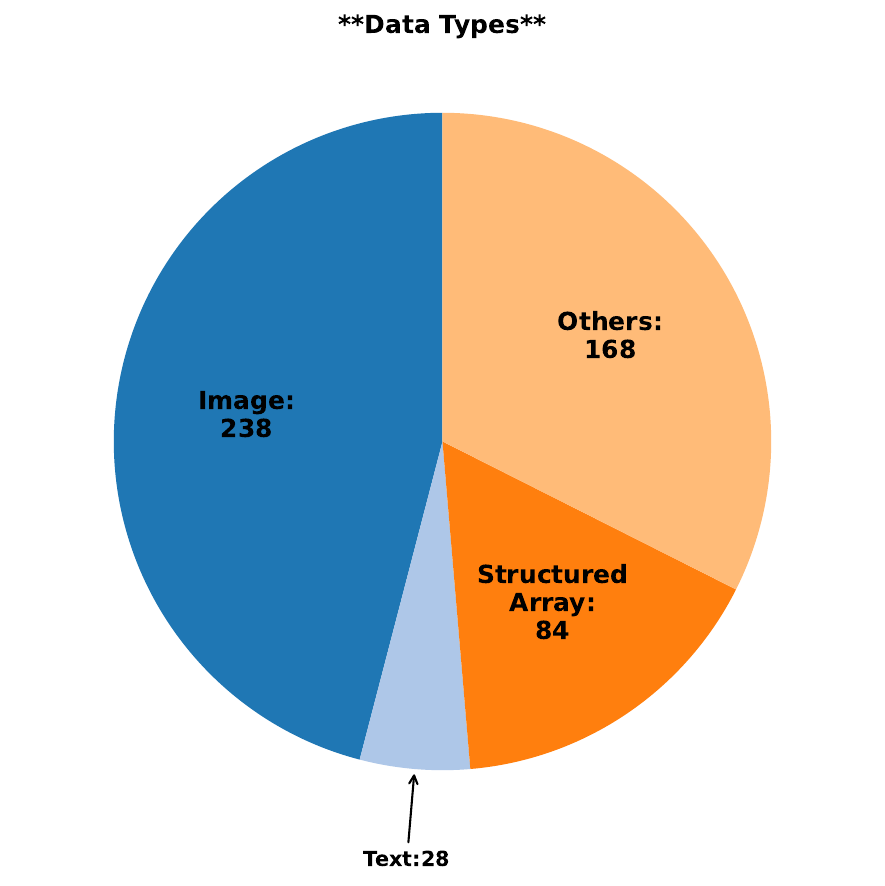}
  }
  \caption{Distribution of code samples in each category}
  \label{fig:statistics}
\end{figure*}

%% file: figs/processing.tex
\begin{figure}[h!]
    \centering
    \includegraphics[width=0.6\linewidth]{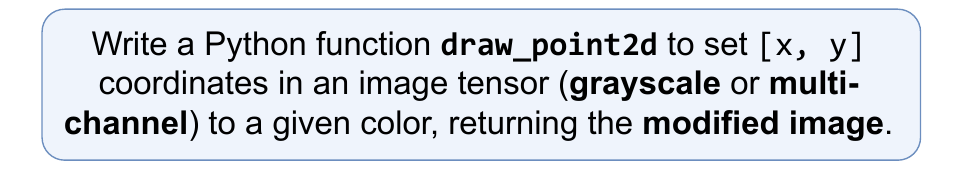}
    \caption{An example prompt 
    for Pre/Post processing
    }
    \label{fig:preproprocessing}
    \vspace{-10pt}
\end{figure}

%% file: figs/model_construction.tex
\begin{figure}[h!]
    \centering
    \includegraphics[width=0.6\linewidth]{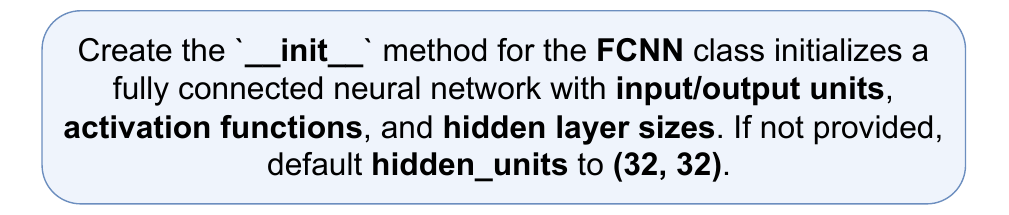}
    \caption{An example prompt 
    for Model Construction
    }
    \label{fig:model_construction}
\end{figure}

%% file: figs/general_example.tex
\begin{figure}[h!]
    \centering
    \includegraphics[width=0.6\linewidth]{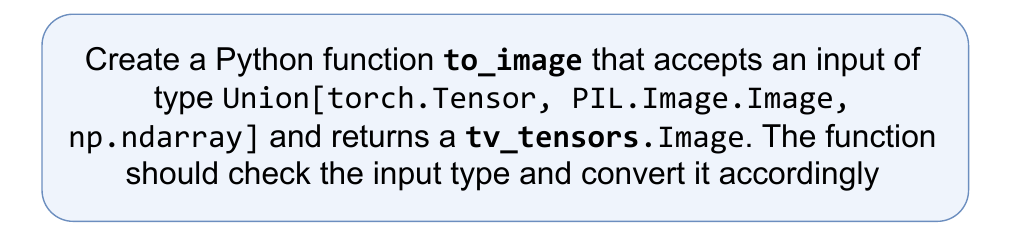}
    \vspace{-7pt}
    \caption{Example of General Task}
    \label{fig:example_general}
    \vspace{-10pt}
\end{figure}

%% file: figs/example_classification.tex
\begin{figure}[h!]
    \centering
    \includegraphics[width=0.6\linewidth]{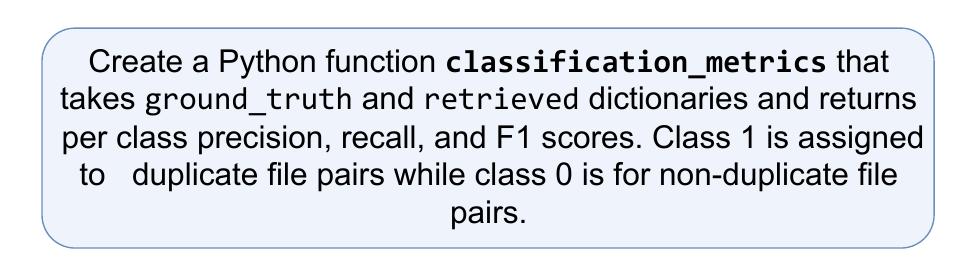}
    \caption{An example of Classification Task.}
    \label{fig:example_classification}
    \vspace{-10pt}
\end{figure}

%% file: tabs/Taxonomies.tex
\begin{table*}[h!]
\centering
     \caption{Distribution of bugs in LLM generated code for deep learning}
    \label{tab:taxonomy}
\resizebox{\linewidth}{!}{
\begin{tabular}{lllc}
\hline
Category &
  DL Related Categories &
  \multicolumn{2}{l}{\# of Occurances} \\ \hline
\multirow{4}{*}{\textbf{Misinterpretation}: Generated code deviates from prompt intention}           & Incorrect DL library or framework Usage& 10& \multirow{4}{*}{120} \\
 &
  Shape and dimension mismatch &
  45 &
   \\
 &
  Incorrect DL/ML Functionality &
  13 &
   \\
 &
  Not DL-related&
  52&
   \\ \hline
\textbf{Syntax Error}: Missing parenthesis, semicolon, or other syntax issues&
   &
   &
  0 \\ \hline
\textbf{Silly Mistake}: Redundant conditions, unnecessary casting&
  Not DL-related &
  8 &
  8 \\ \hline
\textbf{Prompt biased Code}: Code overly relies on examples from the prompt&
  Not DL-related &
  4 &
  4 \\ \hline
\multirow{3}{*}{\textbf{Missing Corner Case}: Edge cases not handled }&
  Tensor Type and Value Edge Cases&
  8&
  \multirow{3}{*}{33} \\
 &
  Shape and Dimension Edge Cases &
  15 &
   \\
 &
  Not DL-related&
  10&
   \\ \hline
\multirow{4}{*}{\textbf{Wrong input type}:Incorrect input type in function calls}&
  Tensor shape mismatch&
  3&
  \multirow{4}{*}{64} \\
 &
  \begin{tabular}[c]{@{}l@{}}Incorrect ML/DL function \\ library arguments\end{tabular} &
  16 &
   \\
 &
  Type mismatch problem &
  23 &
   \\
 &
  Not DL-related&
  22&
   \\ \hline
\multirow{3}{*}{\textbf{Hallucinated Objects}: Nonexistent or undefined objects used}&
  Missing or Undefined DL Modules&
  9&
  \multirow{3}{*}{32} \\
 &
  Incorrect Usage of DL Modules &
  12 &
   \\
 &
  Not DL-related&
  11&
   \\ \hline
\multirow{3}{*}{\textbf{Wrong Attribute}: Incorrect/nonexistent attributes for objects or modules}& Wrong DL Module import& 8& \multirow{3}{*}{46}  \\
 &
  Incorrect API Usage &
  17 &
   \\
 &
  Not DL-related&
  21&
   \\ \hline
\textbf{Non-Prompted Consideration}:Non-requested features added&
  Not DL-related &
  12 &
  12 \\ \hline
\multirow{4}{*}{\textbf{Operation/Calculation Error}:Errors in arithmetic or logical operations} &
  Data Type Casting Issues&
  5&
  \multirow{4}{*}{72} \\
 &
  Shape and Dimension Errors in Operations &
  28 &
   \\
 &
  Incorrect Algebraic Calculations &
  18 &
   \\
 &
  Not DL-related&
  21&
   \\ \hline
   \multirow{2}{*}{\textbf{Performance Issue}: Poor Performance} &
  DL performance issue&
  2&
  \multirow{2}{*}{3} \\
 &
  Not DL-related&
  1&
   \\ \hline
\multirow{2}{*}{\textbf{Prompt missing information}:Incomplete or unclear prompts} &
  Not defining correct dl library&
  4&
  \multirow{2}{*}{10} \\
 &
  Not DL-related&
  6&
   \\ \hline
\textbf{Incorrect or undefined variable/method references}:\\Variables or methods that are not defined or incorrectly referenced&
  Not DL-related &
  11 &
  11 \\ \hline
\textbf{Constant Value Error}:Incorrect constant value assignment &
  Incorrect Tensor Constant Value &
  6 &
  6 \\ \hline

\end{tabular}
}
\end{table*}